\documentclass[twocolumn,showpacs,floatfix,prl]{revtex4}
\usepackage{amssymb}
\usepackage{graphicx}
\usepackage{dcolumn}
\usepackage{bm}
\usepackage[usenames]{color}
\usepackage{ulem}

\begin{document}

\title{Quantum Dynamics of the Hubbard-Holstein Model in Equilibrium
and Non-Equilibrium: Application to Pump-Probe Phenomena}
\author{G.~De~Filippis$^1$, V.~Cataudella$^1$, E. A. Nowadnick$^{2,3}$,
T.~P.~Devereaux$^{2,4}$,  A.~S.~Mishchenko$^{5,6}$ and
N.~Nagaosa$^{5,7}$}
\affiliation{$^1$SPIN-CNR and Dip. di Scienze Fisiche - 
Universit\`{a} di Napoli Federico II - I-80126 Napoli, Italy \\
$^2$ Stanford Institute for Materials and Energy Science, SLAC National 
Accelerator Laboratory, Menlo Park, California 94025, USA \\ 
$^3$ Department of Physics, Stanford University, Stanford
CA 94305, USA \\
$^4$ Geballe Laboratory for Advanced Materials, 
Stanford University, Stanford, California 94305, USA \\
$^5$ Cross-Correlated Materials Research Group, RIKEN Advanced 
Science Institute (ASI),  Wako 351-0198, Japan \\
$^6$RRC ``Kurchatov Institute'',  123182,  Moscow, Russia \\
$^7$Department of Applied Physics, The University of Tokyo,
7-3-1 Hongo, Bunkyo-ku, Tokyo 113, Japan}


\begin{abstract}
The spectral response and physical features of the 2D Hubbard-Holstein model are 
calculated both in equilibrium at zero and low chemical dopings, and after an
ultra short powerful light pulse, in undoped systems. At equilibrium and 
at strong charge-lattice couplings, the optical conductivity 
reveals a 3-peak structure in agreement with experimental observations. 
After an ultra short pulse and at nonzero electron-phonon interaction, 
phonon and spin subsystems oscillate with the phonon period 
$T_{ph} \approx 80$ fs.
The decay time of the phonon oscillations is about $150-200$ fs, similar to the relaxation time
of the charge system. We propose a criterion for observing these oscillations in high $T_c$ compounds:
the time span of the pump light pulse $\tau_{pump}$ has to be shorter 
than the phonon oscillation period $T_{ph}$.   
\end{abstract}

\maketitle

It is accepted that the physics of the high critical temperature (high $T_c$) cuprate superconductors 
is controlled by the interplay of different kinds of 
elementary excitations. The Mott-Hubbard nature of these systems was quickly recognized, and
highlighted the relevance of the electron-electron interaction (EEI) \cite{lee}.
Angle resolved photoemission 
spectroscopy (ARPES) \cite{ARPES} and optical conductivity (OC) \cite{Basov} measurements revealed 
the importance of the electron-phonon interaction \cite{gunn}.

In spite of the general consensus that both EEI and EPI manifest themselves 
in the physical features of these compounds, 
it is still unclear which of these two interactions is responsible for particular properties. 
This difficulty arises because the EEI and EPI have similar spectroscopic energy scales. 
Ultra-fast pump-probe spectroscopy, which probes the time dependence of
the properties of a system after it was displaced from equilibrium by a laser pump pulse, 
is a promising method for disentangling charge, magnetic, and 
lattice degrees of freedom. Indeed, since laser pulses couple to
charge excitations, the characteristic response times of 
the different subsystems are not determined solely by their own characteristic 
energies but are strongly influenced by the strength of their entanglement 
with the perturbed charge density. 

Although pumping in pump-probe experiments on high $T_c$ compounds
\cite{Gedik,Perfetti,Carbone,Pashkin,Giannetti,DalConte,Dodge} 
or any other material (see, e.g., organic charge-transfer crystals
\cite{Okamoto,Iwai,Kawakami}) 
is always done by an ultra-short pulse of light,
there is a significant variation in the experimental setup. 
Conditions that can vary include fluence, 
photon density (total number of photons per site absorbed during the pulse), 
frequency of the pump photons $\omega_{\scriptsize pump}$, 
pulse duration $\tau_{\scriptsize pump}$, time resolution, and, 
of particular importance, the quantity probed after the time delay.
Examples of such quantities are lattice structure and dynamics measured by
ultra-fast electron diffraction \cite{Gedik,Carbone} and electron 
temperature derived from ARPES \cite{Perfetti}.            
However, one of the most informative pump-probe experimental techniques is 
associated with energy- and time-resolved analysis of the electrodynamic 
response, such as reflectivity \cite{Giannetti,DalConte,Okamoto,Iwai,Kawakami},
transmission \cite{Dodge}, or full OC \cite{Pashkin}. 
In particular, pump-probe experiments on high $T_c$ compounds 
\cite{Gedik,Perfetti,Carbone,Pashkin,Giannetti,DalConte,Dodge} 
revealed three basic timescales. 
The charge immediately reacts at times $<50$ fs. 
The phonons which are directly coupled to charges are involved 
in time scale of 100-200 fs, and the energy is gradually dispensed 
 to the whole phonon subsystem (heat reservoirs) 
by anharmonic interactions during further 500-1000 fs. 

In this letter we study the OC of the two dimensional (2D)
Hubbard-Holstein model (HHM) both in
equilibrium and after an ultra-short powerful light pulse. 
We choose this model because it reproduces the equilibrium OC of the doped compounds, where 
experimental studies resolve three peaks \cite{Lupi}:    
two in the 
infrared region around 2000 cm$^{-1}$ (250 meV) and 5000 cm$^{-1}$ (0.6 eV), 
and one in the visible range around 
$1.6 \times 10^4$ cm$^{-1}$ (2 eV).
Calculations in the framework of the t-J-Holstein model reproduce 
the middle peak at 0.6 eV
\cite{Kyung,Fehske,Cappelluti,OCtJPRL,Vidmar09,OCtJPRB} 
and, in more sophisticated approximations, also can distinguish 
the low energy peak at 2000 cm$^{-1}$ 
\cite{OCtJPRL,Vidmar09,OCtJPRB}. However, since this model projects out  
charge fluctuations, the highest peak at 2 eV is simply absent.
Therefore, although the t-J-Holstein model is well suited to describe 
quasistationary states in the presence of a
constant electric field that is not too large\cite{Vidmar11}, since pumping by high-energy
pulses is associated with transitions into the upper Mott-Hubbard
(or charge transfer) band, the HHM is the only model able to describe 
the time evolution of charge, magnetic, and lattice degrees of
freedom in pump-probe experiments.
We are aware of theoretical studies of pump-probe physics in models similar
to the HHM in one dimension (1D) with classical \cite{Yone09}
and quantum \cite{Matsueda} treatment of phonons, and in 2D with a
classical treatment of phonons \cite{Kawakami,Miya09}.   
Here we present the first full quantum treatment of  
thermodynamic equilibrium and time-resolved optical spectroscopy
of an electron-lattice system described by the HHM in 2D.
Such calculations became feasible due to the recently developed double 
phonon cloud method \cite{DPC} which allows lattice 
degrees of freedom in large systems to be treated without uncontrolled approximations. 

The HHM Hamiltonian coupled to an external classical vector 
potential is $H=H_{t}+H_{U}+H_{EPI}$, where
\begin{eqnarray}
H_{t}=-t \sum_{i,\mu,\sigma} 
(e^{iA(\tau)} c^{\dagger}_{i+\mu,\sigma}  c_{i,\sigma} + H.c.), 
\label{eq2}
\end{eqnarray}
\begin{eqnarray}
H_{U}=U  \sum_{i} (n_{i,\uparrow}-\frac{1}{2})(n_{i,\downarrow}-\frac{1}{2}),
\label{eq3}
\end{eqnarray}
\begin{eqnarray}
H_{EPI}= \omega_0 \sum_{i} a^{\dagger}_{i} a_{i}+
g \omega_0 \sum_{i}  (a^{\dagger}_{i}+a_{i}) (1-n_i).
\label{eq4}
\end{eqnarray}
Here $t$ is the hopping amplitude, $c_{i, \sigma}^{\dagger }$ is the fermionic creation operator, 
$\mu$ is a unit vector along the axes of the lattice,  
$a_i^{\dagger }$ creates a phonon at site $i$ with 
frequency $\omega_0$, and $n_i $ is the electron number operator.
The EPI strength is defined by the dimensionless coupling constant 
$\lambda=g^2 \omega_0/4 t$. 
We choose model paremeters typical for 
high $T_c$ materials: $t=0.25$ eV $\cong$ 2000 cm$^{-1}$ and 
$\omega_0 = 0.2t \cong 400$ cm$^{-1}$.
The value of the Hubbard repulsion $U=10t=2.5$ eV yields low-energy
physics very similar to that of the t-J model with 
$J=4t^2/U=800$ cm$^{-1}$.
The time-dependent potential vector is \cite{Matsueda}:
\begin{eqnarray}
A=A_0 e^{-\frac{(\tau-\tau_0)^2}{\tau_{\scriptsize pump}^2}} 
\cos(\omega_{pump}(\tau-\tau_0)),
\end{eqnarray}
and the OC, after the pulse, is given by \cite{fetter}:
\begin{eqnarray}
\sigma(\omega,\tau)= \frac {1} { M \omega} \Im {
\int_0^{\infty} i 
e^ {i (\omega+i\eta) t} \langle \tau | [j(t),j(0)] | \tau \rangle dt }, 
\end{eqnarray}
where 
$| \tau \rangle=T e^ { -i \int_0^{\tau} H(\tau_{1}) d \tau_{1} } | \tau=0 \rangle$, 
$T$ is the time ordering operator, 
$j(t)$ is the current operator in the Heisenberg representation 
along one of the lattice direction axes, 
$\eta$ is a broadening factor taking into account additional
dissipative processes, and $M$ is the number of lattice sites. 
The $| \tau \rangle$ state is obtained through the Lanczos time propagation method \cite{lanczos}.  
In the following we use periodic boundary conditions on 2D lattices \cite{dagotto} 
with $M=10$ for equilibrium and $M=8$ for pump-probe calculations. 
The charge and spin degrees of freedom are evaluated exactly,
whereas the quantum phonons are treated with the double phonon cloud method
\cite{DPC}. 

\begin{figure}[b]
\flushleft
        \includegraphics[scale=0.407]{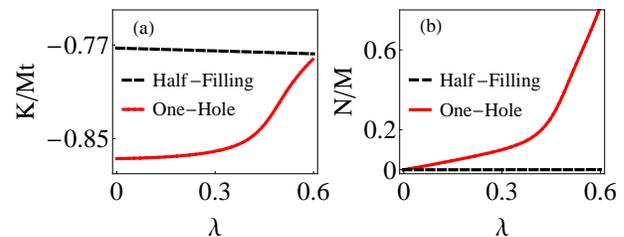}
        \caption{(Color online) (a) Kinetic energy $K$ and (b) 
average number of phonons $N$ versus $\lambda$ at HF and at $\delta=0.1$.}
\label{fig1}
\end{figure}

\begin{figure}[b]
\flushleft
        \includegraphics[scale=0.37]{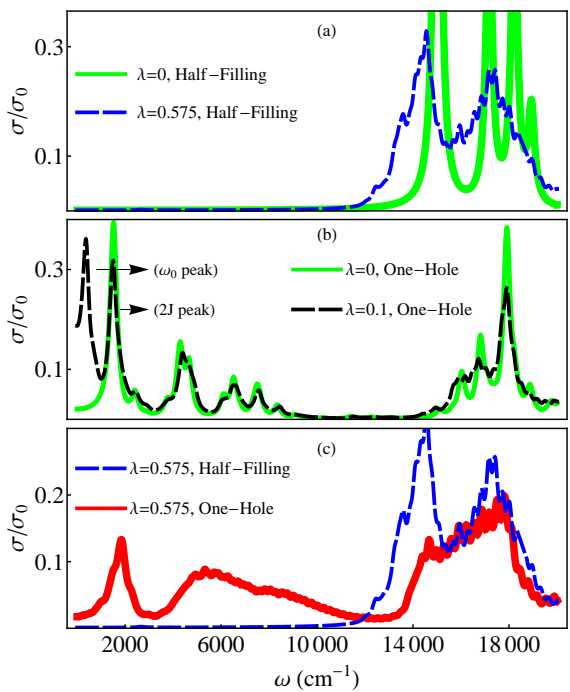}
        \caption{(Color online) OC at equilibrium 
(in units of $\sigma_0=e^2/\hbar a $) for different EPI values, 
at HF and at $\delta=0.1$.} 
\label{fig2}
\end{figure}

\begin{figure}[b]
\flushleft
        \includegraphics[scale=0.40]{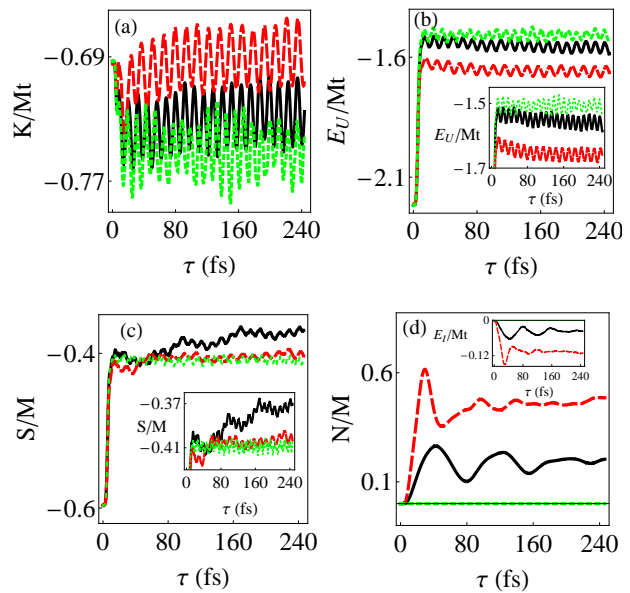}
        \caption{(Color online) Time evolution of:  
(a) Kinetic energy $K$; 
(b) and inset in (b): U interaction contribution $E_U=\langle H_U \rangle$; 
(c) and inset in (c): local spin-spin correlation; 
(d): average number of phonons $N$;
inset in (d): EPI energy at $\lambda=0$ (dotted green line), 
$\lambda=0.1$ (solid black line), and $\lambda=0.5$ 
(dashed red line).
}
\label{fig3}
\end{figure}

\begin{figure}[b]
\flushleft
        \includegraphics[scale=0.37]{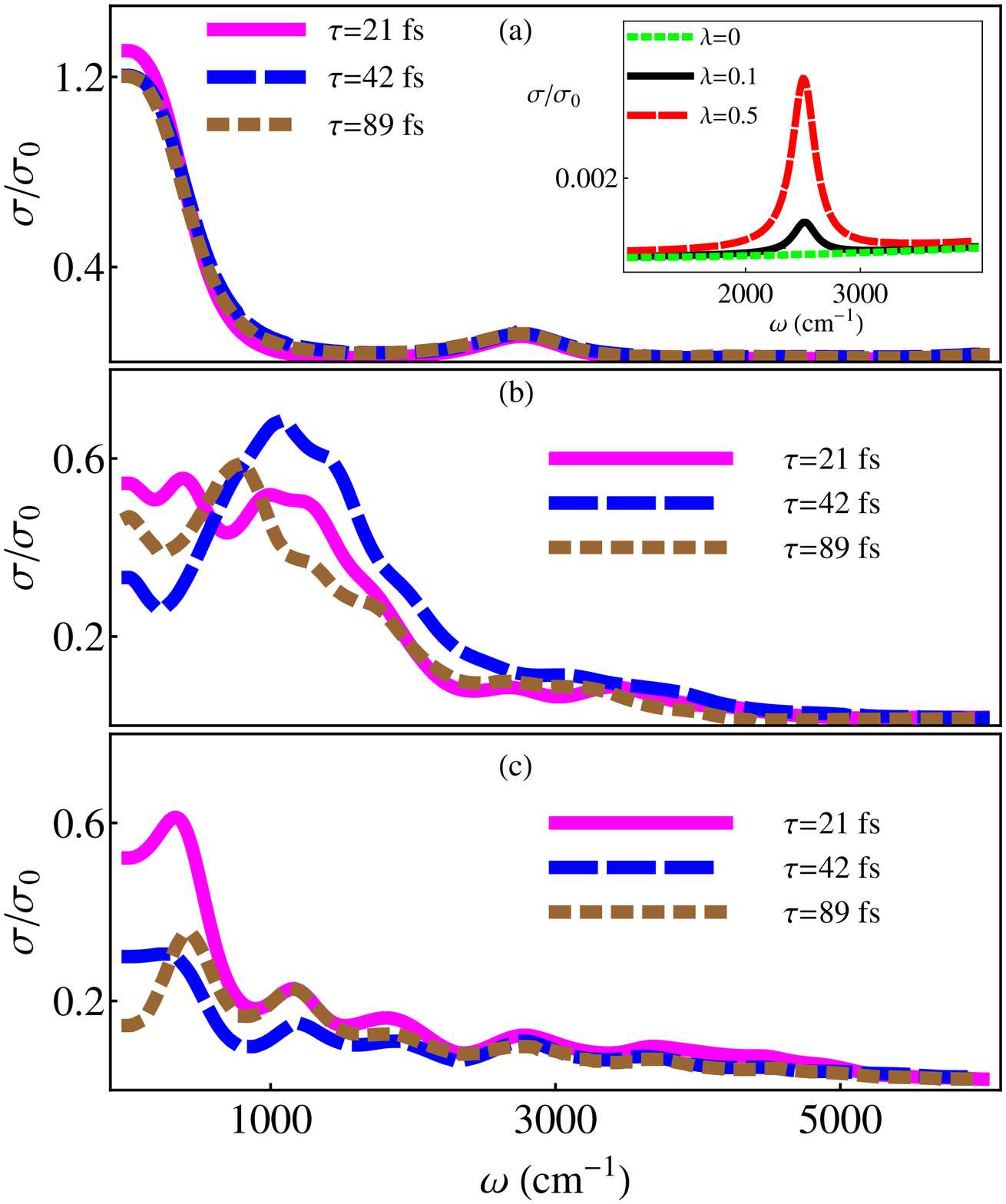}
        \caption{(Color online) Non-equilibrium OC at three different times after 
the pump pulse at (a) $\lambda=0$, (b) $\lambda=0.1$ and (c) $\lambda=0.5$. 
Inset shows zoom of bimagnon feature in the OC at HF and at equilibrium.}
\label{fig4}
\end{figure}

First we study the properties of the HHM in equilibrium.
The kinetic energy $K$ and the average number of phonons 
$N$ dressing the quasiparticle are plotted in Fig.~\ref{fig1}, 
as a function of $\lambda$, at half filling (HF) and at 
hole concentration $\delta=0.1$ away from HF (one extra hole compared to HF 
in a system with $M=10$ lattice sites). The strength of the EPI has a negligible effect 
at HF due to the uniform 
charge distribution, but in the doped system, the doped holes become lattice polarons at  
$\lambda_c \ge 0.4$ \cite{giulio1}.

We now turn to the OC of the  HHM in equilibrium. 
Figure~\ref{fig2}(a) compares the OC at zero ($\lambda=0$) and strong ($\lambda=0.575$) EPI
in the half-filled system; Fig.~\ref{fig2}(b) presents the OC at zero and weak ($\lambda=0.1$) 
EPI at low hole concentration $\delta=0.1$. 
The OC for strong EPI at half filling and at low hole concentration are compared 
in Fig.~\ref{fig2}(c).  
The OC of the HHM at half filling displays, in agreement with experiment \cite{Lupi}, 
a gap with the peak of upper Hubbard band at  
$\omega  \approx 8 t \approx$  16000 cm$^{-1}$.
In contrast to the t-J-Holstein
model, where this upper band is absent, the HHM can describe  
experiments using high-energy pump
photons, $\omega_{pump}>8t$.   
Similar to the ground state properties shown in Fig.~\ref{fig1}, the OC
at HF weakly depends on $\lambda$. 
The high-energy edge of the gap in OC is gradually filled 
by phonon-dressed excitations when $\lambda$
increases.     

On the contrary, the OC of the weakly doped system depends strongly on 
$\lambda$ as shown in Fig.~\ref{fig2}(b).
Upon chemical doping the gap is filled through  
a transfer of spectral weight towards lower energies.   
The most prominent feature of the OC of the weakly doped system 
without EPI is a low energy peak at $\omega \approx 2J = 1600$
cm$^{-1}$ (solid green line in Fig.~\ref{fig2}(b)) which is well known from the 
pure t-J model.
However, this narrow peak cannot explain the broad mid-infrared feature in the underdoped high $T_c$ 
compounds, because its energy is too low, the peak width is too narrow, and, 
for most model parameters, its doping 
dependence is opposite that detected in experiments (see \cite{OCtJPRL,capone,comanac}).   
Besides, the OC without EPI does not display the 2-peak structure
of the experimental mid-infrared response \cite{OCtJPRL}.  
However, even a very small EPI ($\lambda=0.1$) induces a low-energy
peak at $\omega \ge \omega_0$ just above the phonon frequency  
(dashed black line in Fig.~\ref{fig2}(b)). At the same time, the intensity of the
$2J$ peak is renormalized. Finally, the OC at larger values of 
EPI ($\lambda=0.575$) agrees well with 
the experimental positions of the OC peaks at energies 1500-2000 cm$^{-1}$ and  
5000-6000 cm$^{-1}$ (red solid curve in Fig.~\ref{fig2}(c)) observed in strongly underdoped 
high $T_c$ compounds.
The HHM model, like the t-J-Holstein model
\cite{OCtJPRL,EPLKink}, explains the doping dependence of the spectra
in the mid-infrared region. 
In addition, the HHM model describes the experimentally observed 
\cite{Lupi} hardening of the large-energy $\approx$20000 cm$^{-1}$ peak
associated with the charge transfer gap with increasing doping (Fig.~\ref{fig2}(c)).
Recently, it was argued that models with large Coulomb repulsion $U$, leading
to Mott insulators, give OC with too little spectral weight at low energies, and a
smaller value of $U$ must be used in order to agree with experiments \cite{comanac}. 
On the other hand, as shown in Fig.~\ref{fig2}, EPI increases the low energy 
spectral weight of OC and, thus, large $U$ Coulomb repulsion and EPI can provide another 
explanation of the effects discussed in Ref.~[\onlinecite{comanac}]
(see also \cite{CaKoPe}). Time-dependent pump-probe experiments can 
distinguish between these two scenarios.  

We now turn to the study of pump-probe dynamics, by solving  
the time dependent Schroedinger equation. For the chosen hopping $t$, the
real time unit is $\hbar/t = 2.62$ fs.  
We fix $\hbar \omega_{pump}=8.5t$, above the peak
of the charge transfer gap, 
and $A_0=0.2$, corresponding to 
the absorption of about $1/2$ photon during the pulse with 
$\tau_{pump}=2.62$ fs and $\tau_{0}=5.24$ fs. 
We study the dynamics in the first 250 fs because our
model is missing a heat reservoir where energy is dispensed in 
experiments during 500-1000 fs.

The time evolution of the kinetic energy $K$, 
the average value of Hubbard energy $E_U=\langle H_U \rangle$, 
the local spin-spin correlation 
$S=\sum_{i,\mu} \vec{S}_i \cdot \vec{S}_{i+\mu}$, the average number of
phonons $N$, and the average value of the contribution describing
the charge-lattice coupling $E_I=\langle \tau |
g \omega_0 \sum_{i}  (a^{\dagger}_{i}+a_{i}) (1-n_i)| \tau \rangle$
are shown in 
Fig.~\ref{fig3} for three different values of EPI at half filling. 
Quantities associated with charge and spins ($K$, $E_U$, and $S$)
display quick changes during the first 10-20 fs
(Fig.~\ref{fig3}a, b, and c) followed by 
slower dynamics depending on the strength of the EPI.  
On the other hand, quantities associated with lattice degrees of freedom
($N$ and $E_I$) show dynamics on a time scale associated with 
the period of  phonon oscillations 
$T_{ph}=2\pi/\omega_0 \approx 80$ fs (Fig.~\ref{fig3}d). 
Furthermore, after about two phonon oscillations, both at $\lambda=0.1$ and 
$\lambda=0.5$ a relaxed state is reached within a time around 
$150-200$ fs. It is interesting to note that the spin correlation 
function $S$ (Fig.~\ref{fig3}c) 
at $\lambda \ne 0$ also displays oscillations at timescales similar to the
phonon period $T_{ph} \approx 80$ fs, highlighting the 
strong coupling of spin and lattice degrees of freedom. Notably, after 
the laser pulse, slower charge dynamics are observed only in the 
presence of a significant EPI 
(see inset Fig.~\ref{fig3}b): 
at $\lambda=0$, $E_U$ diplays oscillations\cite{note} 
around an essentially constant value, whereas 
at $\lambda=0.5$, a relaxation is observed. The relevance of the 
EPI is also indicated by strong $\lambda$ dependence of $K$ after the pumping: 
the kinetic energy gain, that is $\lambda$ independent in the ground 
state, quickly reduces by increasing $\lambda$\cite{note}.  

The oscillation behavior with characteristic phonon frequencies predicted here,
has not been observed in experiment \cite{Gedik,Perfetti,Carbone,Giannetti,DalConte,Dodge}. 
To our mind, the difference lies in the pump-width to phonon period 
ratio $\tau_{pump} / T_{ph}$ which is small 
$\tau_{pump} / T_{ph} \ll 1$ in our calculation and large
$\tau_{pump} / T_{ph} > 1$ in experiments     
\cite{Gedik,Perfetti,Carbone,Giannetti,DalConte,Dodge}.
This interpretation is supported by numerous pump-probe 
experiments in charge-transfer organic compounds  
\cite{Okamoto,Iwai,Kawakami}, where the relation 
$\tau_{pump} / T_{ph} \ll 1$ holds, and the oscillations of physical 
quantities with phonon period are observed.
Moreover, experiments on high $T_c$ materials with record 
ultra-short $\tau_{pump}=12$ fs pulse \cite{Pashkin} do observe 
oscillations with the frequencies of bond-bending and apical oxygen phonons. 
In addition, the wavelet transformation of time dynamics of the OC shows that 
the energy transfer into the lattice system is associated with a 
time around $\tau_{tr} \approx 150$ fs, which is similar to that of 
phonon oscillations \cite{Pashkin}. 
Hence, interplay of similar periods $\tau_{tr} \approx T_{ph}$ may result 
in beat behavior of lattice properties as observed in our calculations 
at $\lambda=0.5$ (Fig.~\ref{fig3}d).     
  
Fig.\ref{fig4} shows the OC after the pulse at HF 
with (a) zero, (b) weak, and (c) strong EPI. 
The gap present in the undoped system is filled after the pulse, but in a way that is 
strongly dependent on the EPI strength. 
The non-equilibrium OC without EPI shows almost time independent Drude behavior 
(Fig.~\ref{fig4}a). 
On the other hand, for weak EPI ($\lambda=0.1$) as shown in Fig.~\ref{fig4}b, the 
OC oscillates and contains two main peaks at the same energies as observed in the weakly doped system
at weak EPI (Fig.~\ref{fig2}b). These peaks correspond to  $2J$ and $\omega_0$, 
which are related to charge scatterings with magnons and phonons, respectively. These two features, 
associated with spin and lattice degrees of freedom, 
become visible at very short times after the pump pulse. This result agrees with  
results obtained within the Holstein model \cite{Trugi}. 
Finally, the non-equilibrium OC at $\lambda=0.5$ has no  
features in common with the OC at equilibrium (compare Fig.~\ref{fig4}c and Fig.~\ref{fig2}c):
there is no dominant peak at 5000 cm$^{-1}$ or 
at 2000 cm$^{-1}$.  
Instead, a low energy peak with phonon sidebands is observed 
indicating that the lattice polaron is not yet formed. Indeed polaron 
formation at intermediate EPI requires long times\cite{Trugi}, 
due to the high potential barrier between free and self-trapped states.

At zero (weak) EPI, a feature is clearly (tentatively) revealed in the non-equilibrium OC 
at $\omega \simeq 3J$, due to the transition into the bimagnon state.
This excitation has been observed by Perkins et al.~\cite{perkins} in undoped 
copper oxides and, theoretically, can be seen in the equilibrium OC only  
in the presence of EPI \cite{lorenzana} (see inset in Fig.~\ref{fig4}a for equilibrium OC
at different EPIs). 
However, out of equilibrium, due to lower symmetry, such a transition is allowed 
even at $\lambda=0$. 

In conclusion, we studied spectral properties of underdoped high $T_c$ materials
(2D Hubbard-Holstein model with quantum phonons)
in equilibrium and during the first 250 fs following 
excitation by a laser pulse. In particular, 
we found phonon-governed oscillations that 
can be observed in pump-probe experiments provided that the pulse 
time $\tau_{pump}$ is shorter or at least comparable with
the typical phonon times $T_{ph} = 2\pi / \omega_0$. 
We showed that the analysis of OC in and out of equilibrium in the mid-infrared range 
can reveal the EPI strength in cuprates.

ASM acknowledges support of RFBR 10-02-00047a.
NN is supported by MEXT Grand-in-Aid No.20740167, 19048008, 
19048015, 21244053, Strategic International Cooperative
Program (Joint Research Type) from Japan Science and 
Technology Agency, and Japan Society for the 
Promotion of Science through ``Funding Program for
World-Leading Innovative R \& D on Science and Technology 
(FIRST Program)''. TPD and EAN acknowledge support 
from  the U. S. Department of Energy, 
Office of Basic Energy Science, Division of Materials Science and 
Engineering under Contract No. De-AC02-76SF00515.


\begin{thebibliography}{99}
\bibitem{note} In addition to physically meaningful time dynamics finite size
oscillations are observed corresponding to electron
motion from one edge of finite system to another with the time scale
$(\hbar/t)\sqrt{8}\sqrt{2} \approx 10$ fs \cite{Matsueda}.
\end{thebibliography}
\end{document}